\documentclass[draftcls, single, onecolumn, 12 pt]{IEEEtran}
\usepackage{amsfonts, cite}
\usepackage{amssymb}
\usepackage{color}
\usepackage{subfig}
\usepackage{scalefnt}
\usepackage{amsmath}
\usepackage{amsfonts}
\usepackage{colordvi}
\usepackage{verbatim}
\usepackage{graphicx}
\usepackage{latexsym}
\usepackage{times}
\newtheorem{theorem}{Theorem}
\newtheorem{lemma}{Lemma}

\newtheorem{corollary}{Corollary}

\begin{document}

\title{Quantization and Bit Allocation for Channel State Feedback in Relay-Assisted Wireless Networks}

\author{Ehsan~Karamad,~\IEEEmembership{Student Member,~IEEE,}
        Behrouz~Khoshnevis,~\IEEEmembership{Member,~IEEE,}
        and~Raviraj~S.~Adve,~\IEEEmembership{Senior~Member,~IEEE}
\thanks{The authors are with the Edwards S. Rogers Sr.~Department of Electrical and Computer
Engineering, University of Toronto, Toronto, Ontario, Canada M5S 3G4
(email: \{ekaramad,bkhoshnevis,rsadve\}@comm.utoronto.ca).}}

\maketitle \thispagestyle{empty}

\vspace*{-0.5in}

\begin{abstract}
This paper investigates quantization of channel state information (CSI)
and bit allocation across wireless links in a multi-source,
single-relay cooperative cellular network. Our goal is to minimize the
loss in performance, measured as the achievable sum rate, due to
limited-rate quantization of CSI. We develop both a channel
quantization scheme and allocation of limited feedback bits to the
various wireless links. We assume that the quantized CSI is reported to
a central node responsible for optimal resource allocation. We first
derive tight lower and upper bounds on the difference in rates between
the perfect CSI and quantized CSI scenarios. These bounds are then used
to derive an effective quantizer for arbitrary channel distributions.
Next, we use these bounds to optimize the allocation of bits across the
links subject to a budget on total available quantization bits. In
particular, we show that the optimal bit allocation algorithm allocates
more bits to those links in the network that contribute the most to the
sum-rate. Finally, the paper investigates the choice of the central
node; we show that this choice plays a significant role in CSI bits
required to achieve a target performance level.
\end{abstract}

\section{Introduction}
\label{sec:introduction}

It is well established that using relays can significantly improve the
communication capacity and reliability of wireless
networks~\cite{AAZHANG:2003}. Based on approaches suggested
in~\cite{Cover:1979}, the work in~\cite{Laneman:2004}
and~\cite{Kramer:2005} analyzes different relaying strategies such as
the decode-and-forward (DF) and amplify-and-forward (AF) relaying
techniques. Our focus is on DF, wherein the relay must decode and then
re-encode the source data. The potential gains associated with the
relay systems, and cooperative diversity in general, has attracted a
great deal of research into the optimization of the relay-assisted
network performance. For wireless networks, the optimization is mainly
in terms of resource allocation, specifically power and/or bandwidth
allocation, relay routing, and selection of relaying
strategy~\cite{Zhao:2007, Reznik:2004,Elzbieta:2008}.

Although the literature on relay network optimization shows that there
are significant performance improvements to be had, most of the
analysis is based on the crucial assumption that some central node has
exact knowledge of the network-wide channel state information (CSI).
This assumption, however, is impossible to satisfy in a practical
system implementations due to the limited resources available for CSI
training and feedback. As a result, in practice, network-wide CSI is
not known perfectly; only a quantized version of the information may be
available via feedback. Since the performance of resource allocation
algorithms in cooperative relay networks depends heavily on the
availability of CSI, it becomes essential to investigate the
performance of resource allocation schemes under the assumption of
limited-rate quantization of CSI.

This paper takes a step in this direction by investigating the uplink
in a relay-assisted wireless cellular network (much of our analysis can
also be applied to the downlink as well). Our system model considers
multiple sources communicating to a base-station (BS) with the help of
a single relay. We assume that while the receiver end in any link has
perfect CSI of that link, a quantized version of all channels is
available at a central node responsible for resource allocation. The
main goal here is to (i) optimize the quantization of CSI and, (ii)
given a constraint on the number of bits for CSI feedback, allocate
those bits across all links in the network.

The performance of cooperative networks with limited CSI has been
addressed in the available literature for several communication
scenarios. As a general result, it is shown that providing even a few
bits of quantized CSI significantly improves the performance of
cooperative systems~\cite{Shaolei:2010, Xiao:2010, Zhaoxi:2009,
Xueying:2011, Yong:2010, Atia:2008}. The authors
of~\cite{Shaolei:2010}, in particular, investigate optimal temporal
resource allocation between the source and the relay under quantized
CSI using a DF relaying strategy and show that even with a single bit
of feedback provides significant gains in bit error rate. The work
in~\cite{Xiao:2010} shows similar results for diversity gains for both
AF and DF relaying strategies.

The authors of~\cite{Atia:2008} investigate the optimum throughput in a
cooperative network using DF. This work maximizes an upper bound on
overall throughput, thereby deriving a suboptimal resource allocation
scheme. The authors of~\cite{Zhaoxi:2009} investigate optimal relay
selection in multi-relay AF cooperative networks. This work
investigates network performance with quantized and statistical CSI;
again a few CSI bits are shown to provide significant performance
gains.

The works mentioned so far do not specifically investigate how to
quantize CSI or allocate bits. The authors of~\cite{Xueying:2011}
consider a cooperative communication system in a cellular network with
inter-cell interference. The paper adopts zero-forcing beamforming and
finds an approximate expression for the received
signal-to-interference-plus-noise ratio (SINR), based on which bit
allocation is optimized across the network links. The authors also
determine the minimum number of CSI bits that are required to
outperform a non-cooperative network. The authors of~\cite{Yong:2010}
design quantization codebooks for the transmit power vectors in a
single-relay network with DF. For the design, however, they adopt the
Lloyd algorithm with Euclidian-distance as the design metric.


This paper takes a different tack by analyzing the loss due to
quantization in a cooperative wireless network. We start by introducing
the \emph{performance loss} as the loss in the maximum achievable sum
rate due to CSI quantization. To the best of our knowledge, an analysis
of this communication scenario under quantized CSI has not been
presented before. Our analysis includes proposing bounds on the
performance loss and, then, using these bounds to formulate and
optimize quantization schemes for the network-wide CSI. Our main
contributions are:
\begin{itemize}
\item derivation of a tight upper bound on the performance loss due to
    quantized CSI for the sum-rate maximization problem in cellular
    networks;
\item using the upper bound to formulate the optimal CSI quantizer
    design problem. \emph{By using the proposed quantizer}, the
    bound on the performance loss is shown to grow extremely slowly
    with respect to the average link signal-to-noise ratio (SNR)
    and, as is more common, decreases exponentially with the number
    of quantization bits;
\item investigating the optimal allocation of CSI quantization bits across the wireless
    channels to minimize the performance loss. It is shown that most of the quantization bits should
    be used for the links that contribute the most to the sum-rate;
\item a discussion of the choice of the central node to show that
    this choice can have a significant effect on the CSI required
    to achieve a given performance target.
\end{itemize}

The remainder of this paper is organized as follows.
Section~\ref{sec:system model} presents the system and CSI quantization
models. Section~\ref{sec:bound on performance loss} derives an upper
bound on the performance loss due to quantization.
Section~\ref{sec:derivation of the optimal quantizer} then formulates
the optimization problem for quantizer design and presents the
corresponding performance analysis. This is followed by
Section~\ref{sec:central node and bit allocation}, which investigates
optimal bit allocation to minimize the upper bound on the performance
loss due to quantized CSI and discusses the selection of the central
node. Finally, Section~\ref{sec:conclusion} concludes the paper.

\section{System Model}
\label{sec:system model}

The network model comprises $N_S$ source nodes ($S_1 \dots S_{N_S}$)
communicating with a single destination $D$ through a single relay node
$R$. To avoid multiuser interference, each source is allotted an
orthogonal channel. This model most closely represents the uplink of a
relay-assisted cellular network, where due to the unavailability of
direct source-destination ($S_i$-$D$) links, a relay node is deployed
between the source nodes (mobile users) and the destination (base
station) to facilitate communication.

Transmission occurs in two consecutive time slots: the first time slot
is dedicated to source-relay transmission, while in the second time
slot, the relay, using DF, forwards the source messages to the
destination. We further assume that the receiver, in any specific link,
knows the CSI of that specific link exactly, e.g., via adequate
training at the start of each the transmission phase. Such channel
estimation is generally necessary to demodulate and is not an
additional requirement imposed by the resource allocation process.

We assume there exists a central node that collects the quantized
network-wide CSI. This node is responsible for optimal power allocation
at the relay by using the available quantized CSI. Since the relay uses
DF, only the channel magnitudes are required. Specifically, to
calculate the optimal power allocation required at $R$, the central
node needs the magnitudes of the $R$-$D$ and all the $S_i$-$R$
channels.

We assume that the long-term \emph{average channel powers of all links}
are known a priori at the central node. These average powers are
functions of the large scale fading parameters of the links that vary
slowly as compared to the instantaneous channel values. The channel for
a link between a transmitter $X$ and receiver $Y$ is denoted by
$g_{XY}$ and the corresponding normalized channel power is defined as
$h_{XY} = |g_{XY}|^2 / E[|g_{XY}|^2]$. Here, $E[~\cdot~]$ denotes
expectation. Since the average power, $E[|g_{XY}|^2]$, is known at the
central node, we focus on quantizing the normalized channel power,
$h_{XY}$. The probability density functions (pdf) of all the
\emph{normalized} channels are assumed identical for all links. For the
random normalized channel power $h$, $f_H(h)$ and $F_H(h)$ denote,
respectively, the pdf and cumulative distribution function (cdf).
Finally, we assume $f_H(h)$ is bounded and has a bounded derivative
almost everywhere.

The central node is to be given some knowledge of the channel powers of
all the links in the network. In the case of quantized CSI for the link
$X$-$Y$ with $\log_2 N$ bits for quantization, the quantization rule
$q[h_{XY}]$ is implemented as follows:

\begin{itemize}

\item the range $[0, \infty)$ is divided into $N+1$ disjoint
    quantization intervals defined by their boundaries
    $\{q_n\}_{n=-1}^{N}$ where $q_{-1} = 0$ and $q_{N} = \inf \{h
    \ge 0 : F_{H_{XY}}(h) = 1 \}$, i.e., the maximum possible value
    of $h$ (for many pdfs, $q_N = \infty$.). Note that $N$ - and so
    the boundaries (and associated intervals) - may be different
    for different channels.

\item The receiver node, $Y$, observes its instantaneous normalized
    channel power $h_{XY}$ and when this value falls within the
    $n$-th interval, i.e., $h_{XY} \in [q_{n - 1}, q_n)$, the index
    $n$ is fed back to the central node.

\item The central node then assumes the quantized channel power as
    $q[h_{XY}] = q_{n-1}$, i.e., the most conservative value is
    chosen so the resulting sum rate obtained can be guaranteed.
\end{itemize}

On receiving the network-wide CSI, the central node calculates the
power allocation (or equivalently the rate allocation) at the relay
node for all the sources. The relay has a power constraint of $P_R$.
The resource allocation problem for sum-rate maximization is:
\begin{align}
\label{eq:sum_rate_obj}
\max_{\mathbf{P}} \; \; &\sum_{i = 1} ^{N_S} R_i  \\
\label{eq:rel_power_constraint}
\textmd{subject~to:} ~~&\mathbf{1}^T\mathbf{P} \le P_{RD},
\end{align}
where $R_i$ is the rate achieved by source $S_i$ and $\mathbf{1}$ is a
length-$N_s$ vector of ones. In~\eqref{eq:rel_power_constraint},
$P_{RD} = |g_{RD}|^2(P_R/\sigma^2)$ is the SNR at the destination
($\sigma^2$ denotes the noise variance and $P_R$ the power at the
relay). Accounting for the $R$-$D$ channel gain within the power
constraint simplifies the notation in the upcoming analysis. The
optimization is over the vector $\mathbf{P} = \left[ P_1, \dots,
P_{N_S} \right]^T$ which also includes the $R$-$D$ channel gain. $P_i$
then denotes the receive SNR (at the destination) that the relay node
provides to the source node $S_i$. This SNR is the actual power
allocated by the relay to source $S_i$ multiplied by the factor of
$|g_{RD}|^2/\sigma^2$.

Let  $P_{S_iR} = h_{S_iR}(P_S/\sigma^2)$, where $P_S$ is the source
transmit power. Then $R_i$ is given by~\cite{Laneman:2004}:
\begin{align}
\label{eq:rate_source_i}
R_i = \min(C(P_{S_iR}), C(P_i)),
\end{align}
with $C(p) = \ln(1 + p)$, i.e., rate is measured in nats.

To further simplify the notation, we express $P_{RD}$ and $P_{S_iR}$ in
terms of the normalized channel powers, by writing $P_{RD} =
\gamma_{RD} h_{RD}$ and $P_{S_iR} = \gamma_{S_iR} h_{S_iR}$, where
$\gamma_{RD} = (P_R/\sigma^2)E[|g_{RD}|]^2$ and $\gamma_{S_iR} =
(P_S/\sigma^2)E[|g_{S_iR}|^2]$ are the average SNR for the $R$-$D$ and
$S_i$-$R$ links, respectively.

Let $R_i^*$ denote the optimal transmission rate of source $S_i$
obtained by solving~\eqref{eq:sum_rate_obj} assuming perfect CSI.
Similarly, let $R_i^{q*}$ denote the solution to the same problem using
quantized CSI, i.e., the solution to~\eqref{eq:sum_rate_obj} when one
replaces $P_{S_iR}$ and $P_{RD}$ with $q[P_{S_iR}] = \gamma_{S_iR}
q[h_{S_iR}]$ and $q[P_{RD}] = \gamma_{RD} q[h_{RD}]$. Our main goal is
to investigate the performance loss due to quantization, i.e. the
difference between the sum-rate found by
solving~\eqref{eq:sum_rate_obj} with perfect and quantized CSI. We
address this problem in the next section by deriving tight bounds on
the performance loss.

\section{Upper Bound on Performance Loss}
\label{sec:bound on performance loss}

Throughout this paper, the term performance loss or simply loss refers
to the difference between the optimal sum-rate for the perfect and
quantized CSI scenarios. In this section, we provide an upper bound on
this loss in terms of the quantization levels and CSI statistics. This
bound is then used in Section~\ref{sec:derivation of the optimal
quantizer} to optimize the quantizer and eventually derive the optimal
bit allocation across the links in Section~\ref{sec:central node and
bit allocation}. The performance loss is defined as
\begin{align}
\label{eq:general loss all nodes}
\Delta = \sum_{i = 1} ^ {N_S} {\Delta}_i = \sum_{i = 1} ^ {N_S} \left(R_i^* - R_i ^{q*}\right),
\end{align}
where $\Delta_i$ represents the rate loss seen by source $S_i$. We are
interested in the expectation of this loss, i.e., the expected value
of~\eqref{eq:general loss all nodes} over the channel variables. For
each node $i$ define
\begin{align}
\label{eq:general loss node i 1}
E[{\Delta}_i] = E \left[R_i^* - R_i^{q*} \right] =
            E \left[ \min(C(P_{S_iR}), C(P_i^*)) - \min(C(q[P_{S_iR}]), C(P_i^{q*})) \right].
\end{align}
In~\eqref{eq:general loss node i 1}, $P_i^*$ and $P_i^{q*}$ are,
respectively, the optimal power (including the channel gain) allocated
by the relay to source $S_i$ in the perfect CSI and quantized CSI
cases.

Due to the function $\min(\cdot, \cdot)$ in~\eqref{eq:general loss node
i 1} the integration region is divided into four distinct sets. In
order to distinguish these sets, for the source $S_i$, define $A_i =
\left\{ \mathbf{h}: P_{S_iR} \le P_i^* \right\}$ and similarly, $B_i =
\left\{\mathbf{h}: q[P_{S_iR}] \le P_i^{q*} \right\}$. Here,
$\mathbf{h} = \left[h_{S_1R}, h_{S_2R} \dots h_{S_{N_S}R},
h_{RD}\right]^T$ is the vector of variables to be quantized. The sets
$A_i$ and $B_i$ are, respectively, the regions where the source-relay
channel capacity is the bottleneck for the perfect and quantized CSI
scenarios. By definition, the capacity function $C(\cdot)$ is
increasing and~\eqref{eq:general loss node i 1} can be expressed as
\begin{align}
\label{eq:general loss node i 2}
E[{\Delta}_i] &= \int_{\mathbf{h} \in A_i \cap B_i} \left( C(P_{S_iR}) - C(q[P_{S_iR}]) \right)
                        f_{\mathbf{H}}(\mathbf{h}) d\mathbf{h} +
                        \int_{\mathbf{h} \in A_i^c \cap B_i} \left( C(P_{i}^*) - C(q[P_{S_iR}]) \right)
                        f_{\mathbf{H}}(\mathbf{h}) d\mathbf{h}\nonumber \\
    &+ \int_{\mathbf{h} \in A_i \cap B_i^c} \left( C(P_{S_iR}) - C(P_i^{q*}) \right)
        f_{\mathbf{H}}(\mathbf{h}) d\mathbf{h} +
            \int_{\mathbf{h} \in A_i^c \cap B_i^c} \left( C(P_i^*) - C(P_i^{q*}) \right)
            f_{\mathbf{H}}(\mathbf{h})d\mathbf{h},
\end{align}
where $A_i^c$ and $B_i^c$ represent the complements of $A_i$ and $B_i$.
>From the definitions of $A_i$ and $B_i$,
\begin{align}
\label{eq:general loss node i 3}
C(P_{i}^*) \le C(P_{S_iR}) \; \forall \mathbf{h} \in A_{i}^c \cap B_i, \\
\label{eq:general loss node i 4}
C(P_{S_iR}) \le C(P_i^*) \; \forall \mathbf{h} \in A_i \cap B_i^c.
\end{align}
Now from~\eqref{eq:general loss node i 2},~\eqref{eq:general loss node
i 3}, and~\eqref{eq:general loss node i 4} we have the following upper
bound on the performance loss
\begin{align}
\label{eq:general loss node i 5}
E[{\Delta}_i] & \le {\Delta}_{S_iR} + {\Delta}_{RD,i}, \\
\label{eq:general loss node i 6}
\mathrm{where} \hspace*{0.3in} {\Delta}_{S_iR} & = \int_{\mathbf{h} \in B_i}
                                                \left( C(P_{S_iR}) - C(q[P_{S_iR}]) \right)
                                                f_{\mathbf{H}}(\mathbf{h}) d\mathbf{h}, \\
\label{eq:general loss node i 7}
\mathrm{and}\hspace*{0.3in} {\Delta}_{RD, i} & = \int_{\mathbf{h} \in B_i^c}
                    \left( C(P_i^*) - C(P_i^{q*}) \right)f_{\mathbf{H}}(\mathbf{h}) d\mathbf{h}.
\end{align}
Equation~\eqref{eq:general loss node i 6} is an upper bound on the
average performance loss due to quantization of the link $S_i$-$R$ and
is found by merging the first two terms of~\eqref{eq:general loss node
i 2} using~\eqref{eq:general loss node i 3} ;
similarly,~\eqref{eq:general loss node i 7} defines an upper bound the
performance loss due to the power allocated to source $S_i$ based on
quantization of the link $R$-$D$ derived from the third and fourth
terms of~\eqref{eq:general loss node i 2} using~\eqref{eq:general loss
node i 4}.

A similar analysis can be proposed to derive a lower bound
on~\eqref{eq:general loss node i 2} with an expression
resembling~\eqref{eq:general loss node i 5} where the integration
region is replaced by the set $A_i$ and $A_i^c$. Since we are focusing
on the achievable rate regions for the proposed system model, we will
continue with the upper bound on performance loss which ultimately
leads to a lower bound on the achievable rates. In the next section, we
further bound the terms in~\eqref{eq:general loss node i 5}.

\subsection{Loss due to the Quantization of $S_i$-$R$ Links}
\label{sec:S-R Link loss}

In this section we focus on analyzing~\eqref{eq:general loss node i 6}.
From our assumption that all channels are mutually independent, the
joint pdf of $\mathbf{h}$, $f_{\mathbf{H}}(\mathbf{h})$, has a product
form. However, in~\eqref{eq:general loss node i 6}, the region of
integration is coupled across the channel variables making the
integration complicated. To overcome this problem we define a larger
region, $B_i^t$, which includes $B_i$, and results in a product form
for the integral region in~\eqref{eq:general loss node i 6}. The set
$B_i^t$ is defined as
\begin{align}
\label{eq:Bi^t definition}
B_i^t = \left\{\mathbf{h}: q[P_{S_iR}] \le P_{RD}\right\} =
\left\{\mathbf{h}: \gamma_{S_iR} q[h_{S_iR}] \le \gamma_{RD} h_{RD}\right\}.
\end{align}
To see $B_i \subseteq B_i^t$ note that $\forall i$ and $\mathbf{h} \in
B_i$ we have $q[P_{S_iR}] \le P_i^{q*} \le q[P_{RD}] \le P_{RD}
\Rightarrow \mathbf{h} \in B_i^t$.

Since $B_i \subseteq B_i^t$, we achieve an upper bound on the term
${\Delta}_{S_iR}$. From~\eqref{eq:general loss node i 6}
and~\eqref{eq:Bi^t definition}:
\begin{align}
\label{eq:loss SiR Final}
{\Delta}_{S_iR} &\le \int_{\mathbf{h} \in B_i^t} \left(C(P_{S_iR}) - C(q[P_{S_iR}]) \right)
                                                f_{\mathbf{H}}(\mathbf{h}) d\mathbf{h} \nonumber \\
&= \int_{0}^{\infty} \int_{h' \ge \frac{\gamma_{S_iR}}{\gamma_{RD}} q[h]}
                \left( C(\gamma_{S_iR}h) - C(\gamma_{S_iR}q[h]) \right)
                f_{H_{RD}}(h') f_{H_{S_iR}}(h) dh' dh \nonumber \\
&= E \left[ \ln\left( \frac{h_{S_iR} + \gamma_{S_iR}^{-1}}{q[h_{S_iR}] +
            \gamma_{S_iR}^{-1}} \right) \left(1 - F_{H_{RD}}(\alpha_i q[h_{S_iR}])\right)\right],
\end{align}
with $\alpha_i = \gamma_{S_iR}^{-1}\gamma_{RD}$ and using $C(p) =
\ln(1+p)$. The expectation is over $h_{S_iR}$.

The term $1 - F_{H_{RD}}(\alpha_i q[h_{S_iR}])$ in~\eqref{eq:loss SiR
Final} shows that, in general, the quantization of one $S_i$-$R$ link
depends on the distribution of the $R$-$D$ link channel power. This
renders the quantization optimization intractable. We therefore
upper-bound~\eqref{eq:loss SiR Final} by dropping the term $1 -
F_{H_{RD}}(\alpha_i q[h_{S_iR}])$:
\begin{align}
\label{eq:loss SiR Post Final}
{\Delta}_{S_iR} \le E_{h} \left[ \ln\left( \frac{h + \gamma_{S_iR}^{-1}}
                                        {q[h] + \gamma_{S_iR}^{-1}} \right) \right].
\end{align}
The upper-bound in~\eqref{eq:loss SiR Post Final} can be minimized with
respect to the quantization rule $q[\cdot]$. This ultimately leads to
the optimal quantization levels for $h_{S_iR}$, the normalized
$S_i$-$R$ channel power. Crucially, by using~\eqref{eq:loss SiR Post
Final} as the objective function to optimize the quantization, the
quantization levels found for quantization of $S_i$-$R$ link depend on
the statistics the $S_i$-$R$ channel power only. As is shown in the
following sections, adopting the upper bound in~\eqref{eq:loss SiR Post
Final} also leads to separable problems for the optimal quantization
design and bit allocation.

After finding the optimal quantization levels based on~\eqref{eq:loss
SiR Post Final} in Section~\ref{sec:derivation of the optimal
quantizer}, we will return to~\eqref{eq:loss SiR Final} in
Section~\ref{sec:evaluation of loss_SiR} to discuss the optimal bit
allocation across the wireless channels.

\subsection{Loss due to the Quantization of $R$-$D$ Link}
\label{sec:R-D Link loss} The analysis for the $R$-$D$ link follows the
same approach as that of Section~\ref{sec:S-R Link loss}.
Following~\eqref{eq:general loss node i 7} we define the $R$-$D$ loss
component $\Delta_{RD}$ as
\begin{align}
\label{eq:loss R-D definition}
{\Delta}_{RD} = \sum_{i = 1} ^ {N_S} {\Delta}_{RD,i}
= \sum_{i = 1} ^ {N_S} \int_{B_i^c} \left( C(P_i^*) - C(P_i^{q*}) \right)
f_{\mathbf{H}}(\mathbf{h}) d\mathbf{h}.
\end{align}
The problem of evaluating~\eqref{eq:loss R-D definition} for a general
distribution function is intractable. Therefore, similar to the
analysis in the previous section, we extend the region of integration
in~\eqref{eq:loss R-D definition}, resulting in an upper bound on
$\Delta_{RD}$. To this end, define  $B = \cup_{i = 1} ^ {N_S} B_i^c$
which readily yields $B_i^c \subset B$. Since the integrand
in~\eqref{eq:loss R-D definition} is positive, we have
\begin{align}
\Delta_{RD,i} & = \int_{B_i^c} \left( C(P_i^*) - C(P_i^{q*}) \right)
            f_{\mathbf{H}}(\mathbf{h}) d\mathbf{h}
\le \int_{B} \left( C(P_i^*) - C(P_i^{q*}) \right) f_{\mathbf{H}}(\mathbf{h}) d\mathbf{h} \\
\label{eq:loss R-D 1}
& \hspace*{0.3in} \Rightarrow {\Delta}_{RD} \le \int_{B}
                \sum_{i = 1} ^ {N_S} \left( C(P_i^*) - C(P_i^{q*}) \right)
                                                f_{\mathbf{H}}(\mathbf{h}) d\mathbf{h}.
\end{align}
The set of power variables $\left\{ P_i^{q*} \right\}_{i = 1} ^ {N_S}$
represent the optimal power allocation maximizing the sum-rate under
the quantized CSI; therefore
\begin{align}
\label{eq:sum C(P_i^q) inequality}
\sum_{i = 1} ^ {N_S} C(P_i^{q*}) \ge \sum_{i = 1} ^ {N_S}
                            C\left( \frac{q[P_{RD}]}{P_{RD}} P_i^*  \right),
\end{align}
which is true since $\left\{ \frac{q[P_{RD}]}{P_{RD}} P_i^* \right\}_{i
= 1} ^ {N_S}$ is a valid, and likely suboptimal, solution to the max
sum-rate problem satisfying the power constraint. From~\eqref{eq:loss
R-D 1} and~\eqref{eq:sum C(P_i^q) inequality} it follows that
\begin{align}
\label{eq:loss R-D 2}
{\Delta}_{RD} &\le \int_{\mathbf{h} \in B} \sum_{i = 1} ^ {N_S}
            \left[ C(P_i^*) - C\left(\frac{q[P_{RD}]}{P_{RD}} P_i^* \right) \right]
                                        f_{\mathbf{H}}(\mathbf{h}) d\mathbf{h} \nonumber \\
&\le \int_{\mathbf{h} \in B}\sum_{i = 1} ^ {N_S} \ln\left(\frac{1 + P_i^*}
            {1 + \frac{q[P_{RD}]}{P_{RD}}P_i^* } \right) f_{\mathbf{H}}(\mathbf{h}) d\mathbf{h}.
\end{align}
For any channel power $\mathbf{h}\in B$ we have $\sum_{i = 1} ^ {N_S}
P_i^* \le P_{RD}$. On the other hand, the integrand in~\eqref{eq:loss
R-D 2} is a concave function of $P_i^*$ for $P_i \ge 0$ and all $i$.
Using Jensen's inequality, we have
\begin{align}
\label{eq:loss R-D Jensen}
\frac{1}{N_S}\sum_{i = 1} ^ {N_S} \ln \left( \frac{1 + P_i^*}
{1 + \frac{q[P_{RD}]}{P_{RD}} P_i^*}\right) \le
        \ln \left( \frac{1 + \frac{P_{RD}}{N_S}}{1 + \frac{q[P_{RD}]}{N_S}}\right).
\end{align}
Using~\eqref{eq:loss R-D 2} and~\eqref{eq:loss R-D Jensen}, we have
\begin{align}
\label{eq:loss R-D 3}
{\Delta}_{RD} &\le N_{S} \int_{\mathbf{h} \in B}
                \ln \left(\frac{1 + \frac{P_{RD}}{N_S}}{1 + \frac{q[P_{RD}]}{N_S}} \right)
                f_{\mathbf{H}}(\mathbf{h}) d\mathbf{h} = N_S \int_{\mathbf{h} \in B} \ln \left( \frac{\frac{N_S}{\gamma_{RD}} + h_{RD}}
        {\frac{N_S}{\gamma_{RD}} + q[h_{RD}]}\right) f_{\mathbf{H}}(\mathbf{h}) d\mathbf{h}.
\end{align}

In order to proceed with the evaluation of~\eqref{eq:loss R-D 3}, we
need to simplify the definition of set $B$. This can be achieved by
applying the following lemma.

\begin{lemma}
\label{lemma:solution to max sum rate} If the solution to the sum rate
maximization problem in~\eqref{eq:sum_rate_obj} for some channel vector
$\mathbf{h}$ leads to $P_i^* \le P_{S_iR}$ for some source $S_i$, then
the following inequality is valid
\begin{align}
\sum_{i = 1} ^ {N_S} P_{S_iR} \ge \sum_{i = 1} ^ {N_S} P_i^*.
\end{align}
\begin{proof}
See Appendix~\ref{app:proof of lemma solution to max sum rate}.
\end{proof}
\end{lemma}

According to the definition of set $B$, we have $P_i^{q*} <
q[P_{S_iR}]$ for at least one $i$. Then by Lemma~\ref{lemma:solution to
max sum rate}, $\sum_{i = 1} ^ {N_S} q[P_{S_iR}] \ge q[P_{RD}]$. This
leads to an alternative representation of the set $B$ as
\begin{align}
\label{eq:redefining B}
B = \cup_{i = 1} ^ {N_S} B_i^c = \left\{ \mathbf{h}: \exists~i \;
\mathrm{such~that~} \;q[P_{S_iR}] \ge P_i^{q*} \right\}
= \left\{ \mathbf{h}: \sum_{i = 1} ^ {N_S} q[P_{S_iR}] \ge q[P_{RD}] \right\}.
\end{align}
According to~\eqref{eq:redefining B}, the integration region
in~\eqref{eq:loss R-D 3} is defined in terms of the quantized values of
channel powers. Due to the complexity of working with quantized random
variables, we will introduce a slightly larger set $B'$ described by
the true $S_i$-$R$ channel powers. Define $B'$ as
\begin{align}
\label{eq:defining B'}
 B' = \left\{ \mathbf{h}: \sum_{i = 1} ^ {N_S} P_{S_iR} \ge q[P_{RD}] \right\} \supset B.
\end{align}
By defining $Y = \sum_{i = 1} ^ {N_S} P_{S_iR} = \sum_{i = 1} ^ {N_S}
\gamma_{S_iR} h_{S_iR}$ it follows for~\eqref{eq:loss R-D 3} that
\begin{align}
\label{eq:loss R-D 4}
{\Delta}_{RD} &\le N_S \int_{\mathbf{h} \in B} \ln \left( \frac{\frac{N_S}{\gamma_{RD}} + h_{RD}}
                                {\frac{N_S}{\gamma_{RD}} + q[h_{RD}]}\right)
                                    f_{\mathbf{H}}(\mathbf{h}) d\mathbf{h} \nonumber \\
           &\le N_S \int_{\mathbf{h} \in B'} \ln \left( \frac{\frac{N_S}{\gamma_{RD}} + h_{RD}}
                        {\frac{N_S}{\gamma_{RD}} + q[h_{RD}]}\right)
                        f_{\mathbf{H}}(\mathbf{h}) d\mathbf{h}\nonumber \\
           &= N_S \int_{0}^{\infty} \int_{y \ge \gamma_{RD} q[h]} \ln
            \left( \frac{h + N_S \gamma_{RD}^{-1}}{q[h] + N_S \gamma_{RD}^{-1}}\right) f_{H_{RD}}(h)
            f_Y(y) dy dh \nonumber \\
           &= N_S \int_{0}^{\infty} \ln \left( \frac{h + N_S \gamma_{RD}^{-1}}{q[h] +
           N_S \gamma_{RD}^{-1}}\right) (1 - F_Y(\gamma_{RD} q[h]))f_{H_{RD}}(h) dh, \nonumber \\
           & \le N_S \int_{0}^{\infty} \ln \left( \frac{h + N_S \gamma_{RD}^{-1}}{q[h] +
           N_S \gamma_{RD}^{-1}}\right) f_{H_{RD}}(h)dh,
\end{align}
where the first inequality uses $B \subset B'$ and the next
separates out the integral into an integral over $Y$ and and $h_{RD}$.
The next step completes the integration over $Y$. The final inequality
drops the $(1 - F_Y(\gamma_{RD} h))$ term as in the previous section.

This series of steps leaves us with the same objective function as that
of~\eqref{eq:loss SiR Final} used for the optimal quantization problem.
Therefore, based on our analysis, the \emph{same quantization
structure} is optimal for the upper bounds derived for all links across
the network. After investigating the optimal quantizer in the following
section, we return to~\eqref{eq:loss SiR Post Final} and~\eqref{eq:loss
R-D 4} in Section~\ref{sec:evaluation of loss_RD} for the analysis of
the optimal bit allocation.

\section{Derivation of the Optimal Quantizer}
\label{sec:derivation of the optimal quantizer}

The general structure of a quantizer requires quantization intervals
followed by a choice of quantization levels. As described in
Section~\ref{sec:system model}, our approach requires the quantization
level to be the lowest value of the chosen quantization interval. The
quantizer is, therefore, completely characterized by the
\emph{quantization vector} $\mathbf{q} = [q_0, q_1, ..., q_{N - 1}]$,
i.e., a vector comprising the $N$ non-zero quantization levels (note that by definition, $q_{-1} = 0$). Then, according
to the results from the previous section, for both the $S_i$-$R$ and
$R$-$D$ channels, the optimal quantizer is the one which minimizes
$\delta(\mathbf{q})$  where
\begin{align}
\label{eq:loss_general_eq_per_link}
\delta(\mathbf{q}) &= E \left[ \ln\left( \frac{h + \gamma^{-1}}{q[h] + \gamma^{-1}}\right)\right]
= \int_{0}^{\infty} \ln\left( \frac{h + \gamma^{-1}}{q[h] + \gamma^{-1}}\right) f_H(h) dh.
\end{align}
In~\eqref{eq:loss_general_eq_per_link}, $\gamma$ is the average SNR and
equals $\gamma_{S_iR}$ for the case of the $S_i$-$R$ channel and
$\gamma_{RD}$ for the $R$-$D$ channel. Moreover, the expectation
in~\eqref{eq:loss_general_eq_per_link} is with respect to $h$, the
instantaneous channel power of the corresponding link, whose
distribution is that of $h_{S_iR}$ or $h_{RD}$ for the $S_i$-$R$ and
$R$-$D$ links, respectively. For the function $\delta(\mathbf{q})$
in~\eqref{eq:loss_general_eq_per_link} we have
\begin{align}
\label{eq:sum_In}
\delta(\mathbf{q}) = \sum_{n = -1}^{n = N - 1} I_n =
\sum_{n = -1}^{n = N - 1} \int_{q_n}^{q_{n+1}}
        \ln\left(\frac{h + \gamma^{-1}}{q_n + \gamma^{-1}}\right) f_H(h) dh,
\end{align}
i.e., $I_n$ is as the component of the expectation integral over the
interval $[q_n, q_{n + 1})$. Note that, as defined earlier, we set two
fixed quantization levels $q_{-1} = 0$ and $q_N = +\infty$ (or $q_{N} =
\inf \{h \ge 0 : F_{H}(h) = 1 \}$ if the pdf has finite support).

In our model we only consider channels distributions with finite
average power. More specifically, we assume $E[h] = 1$. This assures
$\delta(\mathbf{q}) < \infty$ and consequently, from the continuity of
$f_H(\cdot)$, the function $\delta(\cdot)$  is differentiable with
respect to the quantization levels $q_n$, $0 \le n \le N - 1$.
Therefore, the optimal quantization level $q_n$ satisfies the following
\begin{align}
\label{eq:quant_opt_cond}
\frac{\partial}{\partial q_n} \delta(\mathbf{q}) =
    \frac{\partial}{\partial q_n} \left( I_{n - 1} + I_n \right) = 0.
\end{align}
The following theorem presents the fundamental iterative relation
between the optimal quantization levels and is a key contribution of
this paper.
\begin{theorem}
\label{prop:quantization levels} The quantization levels of the optimal
quantizer minimizing $\delta(\mathbf{q})$
in~\eqref{eq:loss_general_eq_per_link} satisfy
\begin{align}
\label{eq:q_n equation quantizer}
(q_n + \gamma^{-1}) \ln\left(\frac{q_n + \gamma^{-1}}{q_{n - 1} + \gamma^-1} \right) =
\frac{F_H(q_{n + 1}) - F_H(q_n)}{f_H(q_n)}, \hspace*{0.2in} 0 \leq n \leq N-1.
\end{align}
\begin{proof}
Refer to Appendix~\ref{app:proof of relation between quantization levels}.
\end{proof}
\end{theorem}

The quantization levels proposed in Theorem~\ref{prop:quantization levels} are optimal for a variety of distributions,
including the uniform distribution. We first investigate the structure
of the optimal quantizer for the uniform distribution and then extend
the results to more general distributions of channel power.

\subsection{Optimal Quantization for the Uniform Distribution}
In this section we focus on the uniform distribution for the channel
power and present the optimal quantization vector $\mathbf{q}$ which
minimizes $\delta(\mathbf{q})$ in~\eqref{eq:loss_general_eq_per_link}.
For the uniform distribution and from the assumption $E[h] = 1$ we have
$f_H(h) = \frac{1}{2}$ for $0 \le h \le 2$ and $f_H(h) = 0$ for $h >
2$. Since the pdf is a constant, for any $0 \le q_n < q_{n + 1} \le 2$,
we have
\begin{align}
\label{eq:cdf pdf relation unifrom}
F_H(q_{n + 1}) - F_H(q_n) = f_H(q_n) (q_{n + 1} - q_n).
\end{align}
This will simplify the optimality condition proposed in
Theorem~\ref{prop:quantization levels}. Essentially it follows
from~\eqref{eq:q_n equation quantizer} and~\eqref{eq:cdf pdf relation
unifrom} that
\begin{align}
\label{eq:iterative formula for q_n}
(q_n + \gamma ^ {-1}) \ln \left( \frac{q_n + \gamma^{-1}}{q_{n - 1} +
                      \gamma^{-1}} \right) & = q_{n + 1} - q_n, \hspace*{0.3in} 0 \leq n \leq N-1,
\end{align}
where $q_N = \inf \{h \ge 0 : F_{H_{XY}}(h) = 1 \} =2$. Setting $n =
N-1$ and adding $(q_{N-1} + \gamma^{-1})$ to both sides leads to
\begin{align}
\label{eq:q_N_1 for uniform}
(q_{N - 1} + \gamma ^ {-1}) \left( 1 + \ln \left( \frac{q_{N - 1} +
        \gamma^{-1}}{q_{N - 2} + \gamma^{-1}} \right) \right)= 2 + \gamma^{-1}.
\end{align}
Theorem~\ref{prop:structure of uniform quantizer} now specifies all $N$
non-zero quantization levels for the uniform distribution.
\begin{theorem}
\label{prop:structure of uniform quantizer} The $n$-th quantization
level of the optimal quantizer for the uniform distribution is given by
\begin{align}
\label{eq:q_n in terms of r_i}
q_n = \frac{\prod_{i = 0}^{i = n} r_i}{\gamma} - \gamma^{-1}, \; 0 \le n \le N
\end{align}
where $r_i$ is an iterated logarithmic sequence defined as
\begin{align}
\label{eq:iterative formula for r_i}
r_i = 1 + \ln r_{i - 1}, \; 1 \le i \le N.
\end{align}
Finally, the optimal value of $r_0$ satisfies
\begin{align}
\label{eq:finding_r_0}
\prod_{n = 0} ^ {N} r_n = 2\gamma + 1.
\end{align}
\begin{proof}
Refer to Appendix~\ref{app:proof of prop structure of uniform quantizer}.
\end{proof}
\end{theorem}
Note that in the high-SNR regime, we can ignore the $\gamma^{-1}$ term
and we have $r_i = q_{i} / q_{i - 1}, 0\leq i \leq N$, i.e., the ratio
of consecutive quantization levels.

While Theorem~\ref{prop:structure of uniform quantizer} clearly defines
the optimal quantizer for the uniform distribution, this distribution
is impractical. Therefore, in the next section, we extend this
quantizer to a general distribution function. This is an intractable
problem for arbitrary $N$ and we focus on the case of asymptotically
large $N$.

\subsection{Asymptotically Optimal Quantization for General Distributions}
The iterated logarithm in~\eqref{eq:iterative formula for r_i} is a
direct result of~\eqref{eq:cdf pdf relation unifrom} which holds
exactly for the uniform distribution. For a general distribution
function $f_H(h)$,~\eqref{eq:cdf pdf relation unifrom} is the result of
a first-order Taylor-series approximation of $F_H(q_{n + 1})$ at $h =
q_n$. This approximation becomes accurate for large $N$, i.e.,
$F_H(q_{n + 1}) - F_H(q_n) \rightarrow f_H(q_n) \left( q_{n + 1} -
q_{n} \right)$ as $q_{n} \rightarrow q_{n + 1}$. As $N \rightarrow
\infty$, we have $q_{n + 1} - q_n \rightarrow 0, n = 0, 1, \dots, N-2$.
Then the optimality condition for the quantization levels $q_0, q_1,
..., q_{N - 2}$ presented in~\eqref{eq:q_n equation quantizer} is
consistent with that of the uniform quantizer in~\eqref{eq:iterative
formula for q_n}. As a result, the quantizer structure presented
through Theorem~\ref{prop:structure of uniform quantizer} is
asymptotically optimal.

However, this statement is not true for all $n$, specifically, $n=N-1$.
For a general distribution, $q_N = \infty$ and the Taylor series
approximation cannot be applied to the last interval, $[q_{N - 1},
\infty)$. Therefore, for a general distribution, $q_{N - 1}$ remains
unspecified. The value of $q_{N - 1}$ should be chosen such that $I_{N
- 1}$ in~\eqref{eq:sum_In} is small, in turn making $\delta(\mathbf{q})
\rightarrow 0$ with $N \rightarrow \infty$. We call a quantizer
\emph{consistent} if $\delta(\mathbf{q}) \rightarrow 0$ as $N
\rightarrow \infty$. Unfortunately, using~\eqref{eq:q_n in terms of
r_i}-\eqref{eq:finding_r_0} results in a bounded value for $q_{N-1}$
even if $N\rightarrow \infty$, in turn making $I_{N-1}$ always
non-zero. This would, therefore, lead to an inconsistent quantizer. In
this respect,~\eqref{eq:finding_r_0} needs to be modified.

The key is to realize that by using~\eqref{eq:q_n in terms of r_i}
and~\eqref{eq:iterative formula for r_i} finding the proper value of
$q_{N-1}$ is equivalent to choosing an appropriate $r_0$. This choice
is based on the behavior of $f_H(h)$ at large $h$; specifically, the
value of $r_0$ needs to increase with $N$ to guarantee that all $I_{n}$
approach zero for large $N$. This can be achieved by
replacing~\eqref{eq:finding_r_0} with the following
\begin{align}
\label{eq:finding_r_0_general}
\prod_{n = 0} ^ {N} r_n = \kappa_N \gamma + 1,
\end{align}
where the constant $2$ in~\eqref{eq:finding_r_0} is replaced with
$\kappa_N$. Here $\kappa_N$ increases with $N$ to ensure that the
quantizer is consistent -  for large $N$, $h$ is quantized such that
$I_{n}$ becomes sufficiently small $\forall n$. The following theorem
develops an appropriate choice of $\kappa_N$ for a general distribution
function.
\begin{theorem}
\label{theorem: consistency} Consider the proposed quantizer with $r_0$
found from~\eqref{eq:finding_r_0_general}. For a general distribution
function $f_H(h)$ with cdf $F_H(h)$, the quantization loss, defined
in~\eqref{eq:loss_general_eq_per_link}, is bounded by
\begin{align}
\label{eq:upper bound on delta(q) in N}
\delta(\mathbf{q}) \le \mathcal{O}\left( \frac{\ln (\kappa_N^*)}{N}\right),
\end{align}
where $\kappa_N^* = F_H^{-1}(1 - N^{-1})$.
\begin{proof}
Refer to Appendix~\ref{app:proof of theorem consistency}.
\end{proof}
\end{theorem}
Note that, using~\eqref{eq:q_n in terms of r_i}
and~\eqref{eq:finding_r_0_general}, the suggested choice of
$\kappa_N^*$ ensures that the probability of the channel falling in the
final interval, i.e., $Pr\{(H > q_{N-1}\}$, is almost $1/N$.
\begin{corollary}
For any channel with $E[h] = 1$, $\kappa_N^* < N$ and therefore the quantizer is consistent.
\end{corollary}
\begin{corollary}
\label{corollary: Rayleigh} Under Rayleigh fading, i.e., $H \sim
e^{-h}$, and $N \rightarrow \infty$, we have $\kappa_N^* = \ln N$ and
\begin{align}
\delta(\mathbf{q}) \le \mathcal{O}\left( \frac{\ln \ln N}{N}\right).
\end{align}
\end{corollary}
\begin{corollary}
For channel power uniformly distributed in [0,2), for any $N > 1$,
$\kappa^* = 2$ and
\begin{align}
\delta(\mathbf{q}) \le \frac{c}{N}
\end{align}
for some fixed constant $c > 0$.
\end{corollary}

\subsection{Performance loss for High Average SNR}
\label{sec:high snr behavior of quantizer}

In the previous section we considered the asymptotic
$N\rightarrow\infty$ case, but for finite SNR. In this section we
consider the reverse and investigate how the quantizer levels
(equivalently ratios) must change as a function of SNR; specifically we
investigate the high-SNR regime. However, we do not assume that
$N\rightarrow\infty$. We are, therefore, interested in the limiting
behavior of $\delta(\mathbf{q})$ defined
in~\eqref{eq:loss_general_eq_per_link} for the optimal quantizer vector
$\mathbf{q}$ designed using Theorems~\ref{prop:structure of uniform
quantizer} and~\ref{theorem: consistency} when $\gamma \rightarrow
\infty$.

To illustrate the importance of this analysis we first consider the
performance of a fixed quantizer, denoted by $\mathbf{q}'$, for
Rayleigh fading. The quantization levels in $\mathbf{q}'$ do not change
with SNR, specifically $q_0'$ is constant with $\gamma$. Then using
$q_{-1}' = 0$ and the concavity of $\ln(\cdot)$ we have the following
lower bound on $I_{-1}$ (defined in~\eqref{eq:sum_In})
\begin{align}
\label{eq:lower bound high SNR fixed quantization}
I_{-1} = \int_{0}^{q_0'} \ln(\gamma h + 1) f_H(h) dH \ge
    \frac{1}{2} q_0 e^{-q_0} \ln \left(\gamma q_0 + 1\right) \simeq \frac{q_0}{2} \ln \gamma.
\end{align}
From~\eqref{eq:lower bound high SNR fixed quantization} we see that
$\delta(\mathbf{q}') \ge \mathcal{O}(\ln \gamma)$ as $\gamma
\rightarrow \infty$. The loss in sum-rate would, therefore, be at least
$\mathcal{O}(\ln\gamma)$. On the other hand, as $\gamma \rightarrow
\infty$, the overall sum rate is also $\mathcal{O}(\ln \gamma)$.
Therefore for a fixed quantizer, at least a fixed percentage of the
transmission rate is lost due to CSI quantization. For small values of
$N$ this loss becomes quite significant.

The key contribution of this section is to show that, in the limit as
$\gamma \rightarrow \infty$, our proposed quantizer results in a
$\delta(\mathbf{q})$ that grows at a pace much slower than $\ln
\gamma$. As a consequence, the relative rate loss due to CSI
quantization tends to zero as the SNR grows. This is true for a wide
class of channel distributions and even for small values of $N$. The
next theorem summarizes the main results on the high SNR behavior of
$\delta(\mathbf{q})$ for the proposed quantizer.

\begin{theorem}
\label{theorem: high SNR} In the high SNR regime, where $\gamma
\rightarrow \infty$, the quantizer described by
Theorem~\ref{prop:structure of uniform quantizer} leads to a
quantization loss $\delta(\mathbf{q})$ (defined
in~\eqref{eq:loss_general_eq_per_link}) which scales as the $N$-th order
iterated logarithm of the average SNR, $\ln ^{(N + 1)} \gamma$,  where
$\ln^{(0)} x = x$ and $\ln ^{(n)} x$ is defined as
\begin{align}
\label{eq:definition of log n}
\ln^{(n)} x = 1 + \ln\left( \ln^{(n - 1)}x \right), \; n > 0, x \ge 1.
\end{align}
\begin{proof}
Refer to Appendix~\ref{app:proof of theorem high SNR}.
\end{proof}
\end{theorem}

Theorem~\ref{theorem: high SNR} shows that the loss,
$\delta(\mathbf{q})$, approaches infinity extremely slowly - at a rate
of $\mathcal{O}(\ln^{(N+1)} \gamma)$. This result is valid for any
finite valued channel distribution function and as long as $N > 1$, the
\emph{relative} performance loss vanishes with $\gamma \rightarrow
\infty$.

\subsection{Numerical Validation}

\begin{figure}
\includegraphics[scale = 0.6]{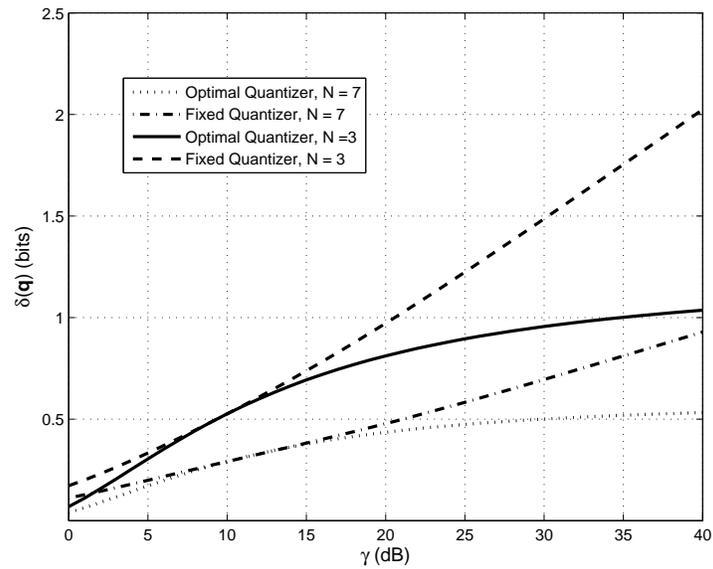}
\centering
\caption{Fixed and optimal quantizers for $N = 3$ ($2$ bits) and $N = 7$ ($3$ bits)
levels of quantization.}
\label{fig:asymptotic_in_SNR}
\end{figure}

To validate our analysis we investigate the loss defined
in~\eqref{eq:loss_general_eq_per_link} through computer simulations.
Fig.~\ref{fig:asymptotic_in_SNR} plots the objective function
$\delta(\mathbf{q})$ in bits (as opposed to nats) for two quantizers:
the first adapts to the average SNR by setting the quantization levels
according to Theorem~\ref{prop:structure of uniform quantizer}. The
second quantizer is similar but the levels are optimized for an average
SNR of $10$dB and then kept fixed. Note that $\delta(\mathbf{q})$ is,
for any specific link, the upper bound on the loss in the rate of that
link - see~\eqref{eq:loss_general_eq_per_link}. This figure is obtained
by numerically generating channel powers drawn from the uniform
distribution  - we use Theorem~\ref{prop:structure of uniform
quantizer} to obtain the quantizer - and then averaging the resulting
$\delta(\mathbf{q})$ over many channel realizations. As
Fig.~\ref{fig:asymptotic_in_SNR} shows, even for $N=3$ there is a
significant difference between the performance of adaptive and fixed
quantizers. Importantly, as suggested in Section~\ref{sec:high snr
behavior of quantizer}, with a fixed quantizer the loss is linear in
average SNR (measured in dB), while, for the optimal quantizer the loss
grows, but very slowly.

\begin{figure}
\includegraphics[scale = 0.6] {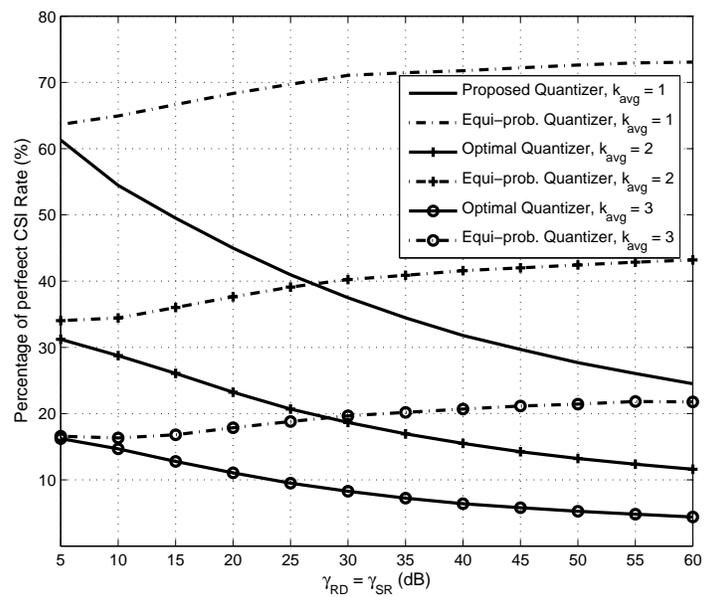}
\centering
\caption{Percentage of the optimal perfect CSI sum-rate lost to quantization.}
\label{fig:loss_ratio_vs_snr}
\end{figure}

In a second test scenario, we simulate a network comprising two source
nodes and a relay node. We assume Rayleigh fading with the same average
SNR for all links. Also all channels deploy the same $k_{avg}$-bit
quantizer ($N +1 = 2^{k_{avg}}$). We compare the optimal quantizer and
the max-entropy quantizer~\cite{Messerschmitt:1971}, i.e., the
quantizer which maximizes the entropy of CSI messages by creating
equi-probable quantization intervals. Fig.~\ref{fig:loss_ratio_vs_snr}
illustrates the performance loss for the two quantizers. The curves in
Fig.~\ref{fig:loss_ratio_vs_snr} show the percentage of the perfect CSI
rate which is lost to quantization as a function of average SNR. As
predicted, this fraction goes to zero for the optimal quantizer while,
for the max-entropy quantizer, it increases as a function of average
SNR and converges (from below) to a constant.

\section{Optimal Bit Allocation}
\label{sec:central node and bit allocation}

In Section~\ref{sec:derivation of the optimal quantizer} we derived the
optimal quantizer based on the upper bound on performance loss
developed in Section~\ref{sec:bound on performance loss}. We then
provided appropriate choices of parameters that ensured a consistent
quantizer. A remaining question is, given a budget on the overall
number of bits available for CSI, how to allocate these CSI bits to
different channels in the network. The key question considered in this
section is how to determine the relative importance of each link in the
network. Based on previous results and assuming the quantizer structure
of Section~\ref{sec:derivation of the optimal quantizer}, we analyze
the performance loss to find the optimal bit allocation algorithm.

To proceed, we continue from the results of~\eqref{eq:general loss node
i 5},~\eqref{eq:loss SiR Final}, and~\eqref{eq:loss R-D 4}. It follows
that
\begin{align}
\label{eq:loss all terms}
{\Delta} &\le \sum_{i = 1} ^ {N_S} E \left[ \ln\left( \frac{h_{S_iR} + \gamma_{S_iR}^{-1}}
                {q_i[h_{S_iR}] + \gamma_{S_iR}^{-1}}\right) (1 - F_H(\alpha_i q_i[h_{S_iR}]))
                \right] \nonumber \\
& \hspace*{1.0in} + N_S E \left[ \ln\left( \frac{h_{RD} + N_S \gamma_{RD}^{-1}}
                {q_{R}[h_{RD}] + N_S \gamma_{RD}^{-1}}\right) (1 - F_Y(\gamma_{RD} q[h_{RD}])) \right],
\end{align}
where we have added indices for the quantizer functions to emphasize
each link is quantized according to different quantization levels.

\begin{figure}
\includegraphics[scale = 0.7]{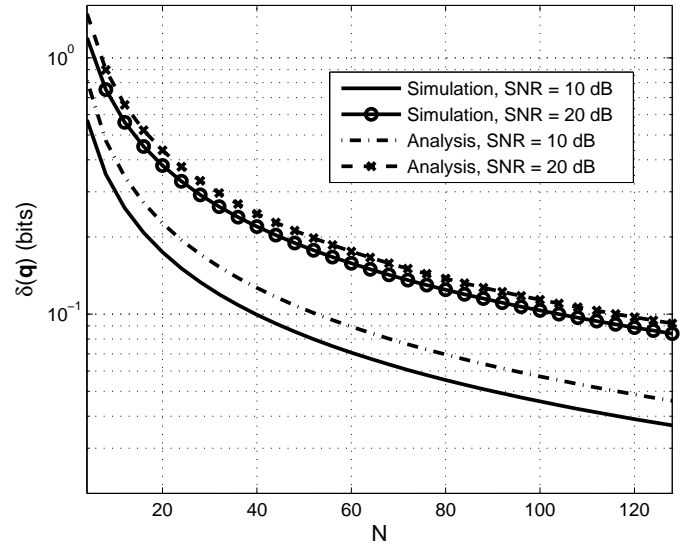}
\centering
\caption{loss vs. number of quantization levels; simulation and analysis for SNR = 10 and 20 dB.}
\label{fig:high resolution bound}
\end{figure}

According to Theorem~\ref{theorem: consistency}, the optimal quantizer
is consistent with $N$, i.e., the performance loss $\delta(\mathbf{q})$
is $\mathcal{O}(\ln(\kappa_N^*)/N)$. Furthermore, the corollaries
showed that, differing distributions of power, $f_H(h)$, result in a
wide variety of appropriate choices of $\kappa_N^*$. However, it
appears that in all cases, the numerator is a relatively slowly
increasing function of $N$, while the denominator is consistently $N$.
To allow for a general - and tractable - analysis, we approximate the
loss as $\delta(\mathbf{q}) \simeq \eta/(N+1)$. This property is tested
in Fig.~\ref{fig:high resolution bound} for the Rayleigh fading channel
distribution ($H \simeq e^{-h})$ where the channel is quantized
according to the proposed quantizer in Section~\ref{sec:derivation of
the optimal quantizer}. The figure shows that the rate of the decrease
in the performance loss is almost $N ^{-1}$.

Following this assumption, ~\eqref{eq:loss all terms} can be written as
\begin{align}
\label{eq:loss all terms approx}
{\Delta} &\le \sum_{i = 1} ^ {N_S} \frac{\eta_i}{2 ^ {k_i}} + \frac{\eta_{RD}}{2 ^ {k_{RD}}}.
\end{align}
where $k_i = \log_2 (N_i+1)$ and $k_{RD} = \log_2 (N_{RD}+1)$ are the
number of CSI quantization bits allocated to the quantization of links
$S_i$-$R$ and $R$-$D$, respectively. This is equivalent to assigning
$N_i$ and $N_{RD}$ quantization levels to the corresponding normalized
channel powers. The coefficients $\eta_i$ and $\eta_{RD}$ are the key
parameters in  defining the bit allocation algorithm across wireless
links and are referred to as the \emph{loss coefficients} for the
$S_i$-$R$ and $R$-$D$ links, respectively. In the next sections we will
extract these loss coefficients based on the results at the end of
Sections~\ref{sec:S-R Link loss} and~\ref{sec:R-D Link loss} and
assuming the optimal quantizer proposed in Section~\ref{sec:derivation
of the optimal quantizer}.

\subsection{Evaluating the loss coefficient for the link $S_i$-$R$: $\eta_i$}
\label{sec:evaluation of loss_SiR} Let $N_i$ be the number of
quantization levels used for the quantization of this link.
From~\eqref{eq:loss SiR Final}
\begin{align}
\label{eq:sum I_n for S_i Redefine}
{\Delta}_{S_iR} \le \sum_{n = -1} ^ {N_i - 1} (1 - F_{H_{RD}}(\alpha_i q_n))
            I_n = I_{-1} + \sum_{n = 0} ^ {N_i - 1} (1 - F_{H_{RD}}(\alpha_i q_n)) I_n,
\end{align}
For large $N_i$ we can eliminate consideration of the edge terms,
setting $I_{-1} , I_{N - 1} \simeq 0$. Also for the optimal quantizer
and for all $q_0 < h < q_{N_i - 1}$, we have $q_n + \gamma_{S_iR}^{-1}
\le h + \gamma_{S_iR}^{-1} < r_{n + 1}(q_n + \gamma_{S_iR}^{-1})$,
which means $\frac{h + \gamma_{S_iR}^{-1}}{r_{n + 1}} - \gamma_{S_iR} ^
{-1} < q_n$ . Therefore, we have
\begin{align}
\label{eq:In upper bound for -1 < n < N - 1}
(1 - F_{H_{RD}}(\alpha_i q_n)) I_n &\le \int_{q_n}^{q_{n + 1}}
\ln \left( \frac{h + \gamma_{S_iR} ^ {-1}}{q_n + \gamma_{S_iR} ^ {-1}}\right)
\left( 1 - F_{H_{RD}}\left(\alpha_i \frac{h}{r_{n + 1}}\right) \right) f_H(h) dh \nonumber \\
&\le \int_{q_n}^{q_{n + 1}} \ln \left( \frac{h + \gamma_{S_iR} ^ {-1}}{q_n + \gamma_{S_iR} ^ {-1}}
\right) \left( 1 - F_{H_{RD}} \left(\alpha_i \frac{h}{r_{1}} \right) \right) f_H(h) dh ,
0 \le n \le N_i - 2,
\end{align}
where the last inequality is true since, from~\eqref{eq:iterative
formula for r_i}, $r_1 \ge r_n \;$ for $n > 1$. Note that the
distribution $f_H(h)$ in this equation corresponds to the $S_i$-$R$
channel under consideration. It follows that
\begin{align}
\label{In upper bound for -1 < n < N - 1 2}
\sum_{n = 0} ^ {N_i - 2} (1 - F_{H_{RD}}(\alpha_i q_n)) I_n &\le
            \int_{q_0} ^ {q_{N_i - 1}} \ln \left( \frac{h + \gamma_{S_iR} ^ {-1}}
                {q_i[h] + \gamma_{S_iR} ^ {-1}}\right)
                \left(1 - F_{H_{RD}} \left(\alpha_i \frac{h}{r_1} \right) \right) f_H(h) dh,
\end{align}
and from the Cauchy-Schwarz inequality we have
\begin{align}
\sum_{n = 0} ^ {N_i - 2} (1 - F_{H_{RD}}(\alpha_i q_n)) I_n &\le \int_{q_0} ^ {q_{N_i - 1}}
        \ln \left( \frac{h + \gamma_{S_iR} ^ {-1}}{q_i[h] + \gamma_{S_iR} ^ {-1}}\right)f_H(h) dh
        \nonumber \\
&\hspace*{0.75in} \times \int_{q_0}^{q_{N_i - 1}} \left( 1 - F_{H_{RD}}
                            \left(\alpha_i \frac{h}{r_1} \right) \right) dh \nonumber \\
&\le \frac{r_1}{\alpha_i} \int_{q_0} ^ {q_{N_i - 1}}
        \ln \left( \frac{h + \gamma_{S_iR} ^ {-1}}{q_i[h] + \gamma_{S_iR} ^ {-1}}\right)f_H(h) dh.
\end{align}
Since $F_{H_{RD}}(\alpha_i q_n)) \le 1$, the coefficient $r_1/\alpha_i$
is replaced by $\min(1, \frac{r_1}{\alpha_i})$ to ensure
that~\eqref{eq:loss SiR Final} does not exceed~\eqref{eq:loss SiR Post
Final}. Finally, from~\eqref{eq:sum I_n for S_i Redefine} and~\eqref{In
upper bound for -1 < n < N - 1 2} we have
\begin{align}
\sum_{n = -1} ^ {N_i - 1} (1 - F_{H_{RD}}(\alpha_i q_n)) I_n &\le I_{-1} +
        \min\left(1, \frac{r_1}{\alpha_i}\right) \sum_{n = 0} ^ {N_i - 2} I_n + (1 - F_{H_{RD}}(\alpha_i q_{N_i - 1})) I_{N_i - 1} \nonumber \\
&\simeq \min (1, \frac{r_1}{\alpha_i}) \sum_{n = -1} ^ {N_i - 1} I_n
= \min \left(1, \frac{r_1}{\alpha_i}\right)
E\left[ \ln \left( \frac{h + \gamma_{S_iR} ^ {-1}}{q_i[h] + \gamma_{S_iR} ^ {-1}}\right) \right].
\end{align}
Based on Theorem~\ref{theorem: consistency} we know that the
expectation above is inversely proportion to $N$. Therefore,
\begin{align}
\label{eq:loss SiR approximation simplified}
{\Delta}_{S_iR} \le \min \left(1, \frac{r_1}{\alpha_i}\right) \frac{c_q}{N_i + 1} =
\min \left(1, \frac{r_1}{\alpha_i}\right) \frac{c_q}{2^{k_i}}.
\end{align}
From~\eqref{eq:loss SiR approximation simplified} we obtain $\eta_i =
\min \left(1, \frac{r_1}{\alpha_i}\right) c_q$ where $c_q$ is a
constant independent of $N_i$.

\subsection{Evaluating the $R$-$D$ loss coefficient: $\eta_{RD}$}
\label{sec:evaluation of loss_RD} For the $R$-$D$ link we follow the
result in~\eqref{eq:loss R-D 4}. Evaluating~\eqref{eq:loss R-D 4} for a
general distribution function $f_H(h)$ is intractable. Therefore, we
resort to the Rayleigh fading channel model leading to the negative
exponential distribution on the channel power.

To find~\eqref{eq:loss R-D 4} we need the cdf function $F_Y(\cdot)$
where $Y$ is a weighted sum of negative exponential random variables as
defined in Section~\ref{sec:R-D Link loss}. We can approximate this
random variable with an Erlang-$2N_S$ random variable. More
specifically, we define the following\footnote{Note that this
approximation results in a tight upper bound on~\eqref{eq:loss R-D 4}.
The intuition is that the random variable $Z$ (with $E[Z] = E[Y_{norm}]
= N_S$) with $Y_{norm} = Y/E[Y]$ has a much smaller variance than
$Y_{norm}$. Therefore, one could imagine the distribution of $Z$ being
more concentrated around its mean, $N_S$, whereas the distribution of
$Y_{norm}$ has a wider spread around its mean. This makes $1 - F_Z(h)
\ge 1 - F_{Y_{norm.}}(h)$ for $h \le N_S$. In the region $h > N_S$, we
already have $f_H(h) = e^{-N_S} \ll 1$ which diminishes the effect of
the approximation error in the overall value of the integral.}
\begin{align}
\label{eq:definition of Z}
Z = 2\sum_{i = 1} ^ {N_S} H_{S_iR} \simeq  \frac{2Y}{\sum_{i = 1} ^ {N_S} \gamma_{S_iR}}.
\end{align}
The random variable $Z$ is a standard Erlang-$2N_S$ with cdf $F_Z(z) =
\left[1 - \sum_{n = 0} ^ {2N_S - 1} e^{-z} \frac{z ^ n}{n!}\right]$.
From~\eqref{eq:definition of Z} we can write~\eqref{eq:loss R-D 4} as
\begin{align}
\label{eq:loss R-D bound with Erlang 1}
{\Delta}_{RD} \le N_S \int_{0}^{\infty} \ln \left( \frac{h + N_S \gamma_{RD}^{-1}}
                {q[h] + N_S \gamma_{RD}^{-1}}\right) (1 - F_Z(\beta q[h]))f_H(h)dh,
\end{align}
where $\beta = 2\gamma_{RD}/\sum_{i = 1} ^ {N_S} \gamma_{S_iR}$. From
the cdf of the Erlang-$2N_S$ distribution,~\eqref{eq:loss R-D bound
with Erlang 1} implies
\begin{align}
\nonumber
{\Delta}_{RD} \le N_S \sum_{k = 0} ^ {2N_S - 1} J_k,
\end{align}
with $J_k$ defined as
\begin{align}
\label{eq:defining J_k}
J_k &=\int_{0}^{\infty} \ln \left( \frac{h + N_S \gamma_{RD}^{-1}}{q[h] + N_S \gamma_{RD}^{-1}}\right)
        \frac{(\beta q[h]) ^ k}{k!} e^{-\beta q[h]} f_H(h) dh =
        \frac{(-\beta) ^ k}{k!} \frac{\partial^k J_0}{\partial \beta ^k}.
\end{align}
Here we assume high resolution quantization such that $\beta q[h] + h
\simeq (\beta + 1) h$ and also $(1 + \beta) q[\frac{h}{1 + \beta}]
\simeq h$. Now by defining the auxiliary variable $u = (\beta + 1) h$,
$J_0$ reduces to the following
\begin{align}
J_0 &\simeq \frac{1}{\beta + 1} \int_{0}^{\infty} \ln \left( \frac{u + N_S(\beta + 1)
\gamma_{RD}^{-1}}{q[u] + N_S (\beta + 1) \gamma_{RD}^{-1}}\right) e^{-u} du \nonumber \\
 &= N_S \frac{1}{\beta + 1} E\left[\ln \left( \frac{h + \gamma^{-1}}{q[h] +
                \gamma^{-1}}\right)\right], \nonumber
\end{align}
with $\gamma ^ {-1} = N_S \left((2\sum_{i = 1}^{N_S} \gamma_{S_iR}) ^
{-1} + \gamma_{RD} ^ {-1}\right)$. On the other hand, from
~\eqref{eq:defining J_k} we have:
\begin{align}
J_k \simeq \frac{\beta ^ k}{(\beta + 1) ^ {k + 1}} E\left[\ln \left( \frac{h + \gamma^{-1}}
                                                        {q[h] + \gamma^{-1}}\right)\right], \nonumber
\end{align}
resulting in
\begin{align}
\label{eq:loss R-D final approximation}
{\Delta}_{RD} \le \sum_{k = 0} ^ {2N_S - 1} J_k = N_S \left(1 - \left(\frac{\beta}
{\beta + 1}\right) ^ {2N_S} \right) E\left[\ln \left( \frac{h + \gamma^{-1}}
{q[h] + \gamma^{-1}}\right)\right].
\end{align}
Similar to the discussion at the end of section~\ref{sec:evaluation of
loss_SiR} we have
\begin{align}
\label{eq:loss R-D final approximation simplified}
{\Delta}_{RD} &\le \sum_{k = 0} ^ {2N_S - 1} J_k =
    N_S \left(1 - \left(\frac{\beta}{\beta + 1}\right) ^ {2N_S} \right) \frac{c_q}{N_{RD}}
= N_S \left(1 - \left(\frac{\beta}{\beta + 1}\right) ^ {2N_S} \right)
           \frac{c_q}{2 ^ {k_{RD}}}.
\end{align}
The result in~\eqref{eq:loss R-D final approximation simplified}
suggests that $\eta_{RD} = N_S \left(1 - \left(\frac{\beta}{\beta +
1}\right) ^ {2N_S} \right) c_q$. Note that since $c_q$ is common to
both the $S_i$-$R$ and $R$-$D$ channels, it is irrelevant.

\subsection{Bit Allocation}
\label{sec:bit allocation} As explained in the beginning of
Section~\ref{sec:central node and bit allocation}, the upper bound
in~\eqref{eq:loss all terms approx} can be used to formulate the bit
allocation problem. In particular, we look at the problem of bit
allocation in a scenario where the system imposes a cap on the overall
number of CSI bits in each transmission phase. We assume this number is
$k_{max} > N_S + 1$. Based on this model and from~\eqref{eq:loss all
terms approx}, the optimal bit allocation problem is formulated as
\begin{align}
\min_{\mathbf{k}} &\sum_{i = 1} ^ {N_S} \frac{\eta_i}{2 ^ {k_i}} +
\frac{\eta_{RD}}{2 ^ {k_{RD}}}  \nonumber \\
subject \; to: &\sum_{i = 1} ^ {N_S} k_{i} + k_{RD} \le k_{max}, \nonumber \\
&k_{i} \ge 1, \; \forall i,
\end{align}
where $k_i$ denotes the number of CSI bits allocated to the $S_i$-$R$
link and $k_{RD}$ the number allocated to the $R$-$D$ link. Finally,
$\mathbf{k} = [k_1, \dots, k_{N_S}, k_{RD}]$. The last constraint
ensures that the transmission rate for all nodes is non-zero (if CSI
bits are not allocated to a link, the corresponding channel power is
quantized to zero; in turn the achievable rate for that link is zero).

The solution to the bit allocation problem follows a simple iterative
algorithm. Initialize the allocation vector as $\mathbf{k}^1$ as the
all-ones vector. Assume $\mathbf{k}^j = [k_1^j, k_2^j, \dots, k_{N_S}^j
k_{RD}^j]$ denotes the pattern of bit allocation at iteration $j$. At the iteration $j + 1$ we look for the link with the
largest effect on the performance loss. Link $n$ has the largest
contribution to the upper bound on loss if $n = \arg \max_{m} 2^{-k_m^j}
\eta_m$. Then, the next bit is allocated to link $n$, i.e., $k_n^{j +
1} = k_n^j + 1$. This procedure is repeated until all $k_{max}$ bits
are allocated to the source nodes (in $k_{max} - N_S - 1$ iterations).

\begin{figure}
\includegraphics[scale = 0.6] {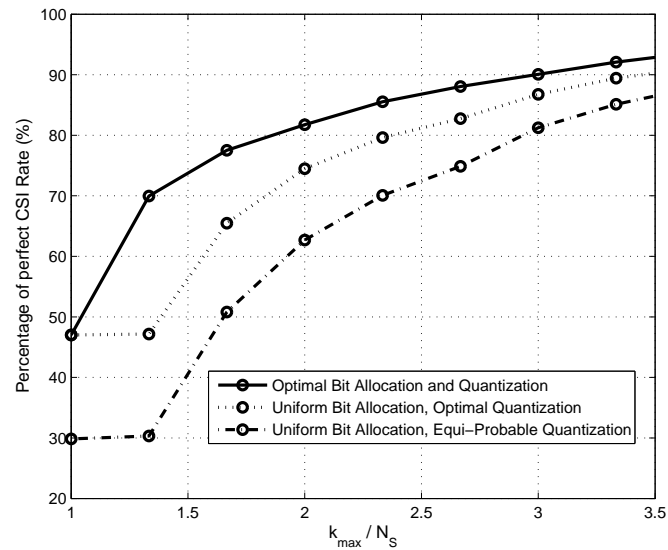}
\centering
\caption{Percentage of the optimal perfect CSI sum-rate achieved through different
methods of quantization and bit allocation for two users.}
\label{fig:bit allocation uniform vs optimal}
\end{figure}

To illustrate the performance gain through optimal bit allocation, we
simulate a two-source network where the sources are randomly located in
front of a relay which is at a fixed distance to the destination. We
assume a Rayleigh fading model for all links and $E[\gamma_{S_iR}]$ =
25dB and $\gamma_{RD}$ fixed at 20dB, i.e., the scenario where the
relay is located closer to the source nodes rather than the
destination. We compare optimal bit allocation and uniform bit
allocation, (i.e., $N_{S_iR} = N_{RD}$). Also we include the
performance of both the max-entropy quantizer and the quantizer
proposed in Section~\ref{sec:derivation of the optimal quantizer} in
order to illustrate the performance gains through optimal quantization.
Fig.~\ref{fig:bit allocation uniform vs optimal} illustrates the
percentage of the perfect CSI sum rate achieved for each case under
quantized CSI. It is seen that through optimal bit allocation we can
achieve considerable performance gain (as opposed to uniform bit
allocation) and this difference is particularly interesting when the
$k_{max}$ is small. Furthermore, Fig.~\ref{fig:bit allocation uniform
vs optimal} shows that our proposed quantizer always outperforms the
max-entropy quantizer. For the given network parameters and at the
$80\%$ target level, it is observed that the proposed quantizer saves
almost one bit per link compared to the max-entropy quantizer while
this saving grows to more than 1.5 bits per link after bit allocation.
One important reason for this difference is the adaptability of
quantization to SNR.

\subsection{Central Node}
In the discussion so far, we assumed that the quantized CSI is reported
to a node that is not part of the network. A more detailed look at the
terms in~\eqref{eq:loss all terms approx} reveals that choice of the
central node will drastically affect the value of performance loss.
Before the uplink data transmission phase starts, the CSI for links
$S_i$-$R$ are already available at the relay and the CSI for $R$-$D$
link is available at the destination via training. Selecting the relay
or the destination as the central node is equivalent to assuming the
perfect CSI of the $S_i$-$R$ links or the $R$-$D$ link is available at
the central node. This is equivalent to letting $k_i$ for all $i$ or
$k_{RD}$ grow to infinity where $k_i$ and $k_{RD}$ are defined
in~\eqref{eq:loss all terms}. In short, the selection between the
destination and the relay reduces to comparing the first and second
terms in~\eqref{eq:loss all terms}.

For the proposed system model it could be argued that relay is the best
choice to serve as the central node. The reason is that each link in
the network needs at least one bit for the CSI quantization (otherwise
the link is assumed to be dead and the channel is always quantized to
zero). This demands a minimum of $N_S$ CSI bits for the quantization of
$S_i$-$R$ links which might lead to large CSI quantization costs for
multiple node networks. The selection of the relay as the central node
will cancel this requirement. At the same time, providing a few bits to
the quantization of the $R$-$D$ link ensures a considerably small
performance loss.

Although setting the relay as the central node is beneficial in terms
of CSI demand, we should mention that for the multiple relay networks,
it is more reasonable to choose the destination as the central node.
This is due to the fact that the destination, as the central node, can
resolve the source-relay assignment problem across the network.

\section{Conclusions}
\label{sec:conclusion} In this paper we developed bounds on the
performance loss due to quantization of CSI in a multi-source,
single-relay, network. Our system model is most similar to the uplink
of a cellular system. Our design metric is the sum-rate achieved over
all the nodes; the relay allocates its power among the source nodes in
order to maximize the sum rate. Our analysis leads to a tight upper
bound on the performance loss which is expanded as the sum of
individual terms each representing the loss due to the quantization of
a certain wireless link.

We use the upper bound on the performance loss to develop an optimal
quantizer. This quantizer is consistent in the sense that the loss
approaches zero as the number of quantization levels increases.
Moreover, one key result we develop is that the quantizer is strongly
robust to the average SNR of the link; the loss is an $N$-th order iterated
logarithm of the average SNR. A consequence is that the performance
loss stays almost constant over the range of practical values for the
average SNR and the relative loss goes to zero as SNR increases.

Using the proposed upper bound and considering the optimal quantizer,
the performance loss is further reduced through optimal bit allocation
across the wireless links. A key contribution is to quantize the
relative importance of each link in the network. Numerical results show
that through quantization and bit allocation, considerable savings in
the average number of CSI bits per node is obtained. Finally, we argue
that for the proposed network model when the number of source nodes is
large, it is better to select the relay as the central node.

\appendices
\section{Proof of Lemma~\ref{lemma:solution to max sum rate}}
\label{app:proof of lemma solution to max sum rate}

We prove the lemma by contradiction. The optimal power allocation is
the solution to~\eqref{eq:sum_rate_obj} with the constraint
in~\eqref{eq:rel_power_constraint}. From~\eqref{eq:rate_source_i},
$R_i$ is increasing in $P_i$, but for $P_i > P_{S_iR}$, $R_i$ remains
constant. Now if there is at least one source node such that $P_i <
P_{S_iR}$, then the optimal power allocation mandates that for all
other nodes we must have $P_j \le P_{S_jR}$. To see this, assume the
opposite, i.e., there is at least one other node $j$ such that $P_j >
P_{S_jR}$. Then, since $R_j$ is constant in a neighborhood around
$P_j$, we can simply reduce $P_j$ by some $\delta P \le P_j - P_{S_jR}$
which maintains the value of $R_j$, and add it to $P_i$ which increases
$R_i$ by $C(\min(P_i + \delta P, P_{S_iR}) - C(P_i)$. Therefore, this
allocation cannot be optimal.

\section{Proof of Theorem~\ref{prop:quantization levels}}
\label{app:proof of relation between quantization levels} The theorem
sets the iterative relationship between the optimal quantization
levels, From the definition in~\eqref{eq:sum_In} we obtain
\begin{align}
\label{eq:expand_I_n + I_(n - 1)}
I_{n - 1} + I_n &= \int_{q_{n - 1}}^{q_{n}} \ln\left(\frac{h + \gamma^{-1}}
            {q_{n-1} + \gamma^{-1}}\right) f_H(h) dh + \int_{q_n}^{q_{n+1}}
            \ln\left(\frac{h + \gamma^{-1}}{q_n + \gamma^{-1}}\right) f_H(h) dh \nonumber \\
&= \int_{q_{n - 1}}^{q_{n + 1}} \ln(h + \gamma^{-1}) f_H(h)dh  \nonumber \\
&- \left(\ln(q_{n - 1} + \gamma ^ {-1}) \int_{q_{n - 1}}^{q_n} f_H(h)dh +
            \ln(q_{n} + \gamma ^ {-1}) \int_{q_n}^{q_{n + 1}} f_H(h)dh \right).
\end{align}
Now from the optimality condition in~\eqref{eq:quant_opt_cond} we have
\begin{align}
\label{eq:proof quantization levels part 1}
\frac{\partial}{\partial q_n} \left( I_{n - 1} + I_n\right)
        &= \frac{\partial}{\partial q_n} \int_{q_{n - 1}}^{q_{n + 1}}
                                \ln(h + \gamma^{-1}) f_H(h)dh
        - \frac{\partial}{\partial q_n} \left(\ln(q_{n - 1} + \gamma ^ {-1})
        \int_{q_{n - 1}}^{q_n} f_H(h)dh \right. \nonumber \\
        & \hspace*{2in}\left.  +\ln(q_{n} +
        \gamma ^ {-1}) \int_{q_n}^{q_{n + 1}} f_H(h)dh \right) = 0.
\end{align}
The first term of~\eqref{eq:proof quantization levels part 1} is
independent of $q_n$, hence the derivative is zero. From the second
term it follows
\begin{align}
\label{eq:proof quantization levels part 2}
&\frac{\partial}{\partial q_n} \left(\ln(q_{n - 1} + \gamma ^ {-1})
\int_{q_{n - 1}}^{q_n} f_H(h)dh + \ln(q_{n} +
\gamma ^ {-1}) \int_{q_n}^{q_{n + 1}} f_H(h)dh \right) \nonumber \\
&\hspace*{0.5in} = \ln(q_{n - 1} + \gamma ^ {-1}) f_H(q_n) +
        \ln(q_{n} + \gamma^{-1}) (-f_H(q_n)) + \frac{\int_{q_n}^{q_{n + 1}} f_H(h)dh}
                        {q_n + \gamma^{-1}} \nonumber \\
&= -f_H(h) \ln\left( \frac{q_n + \gamma^{-1}}{q_{n - 1} + \gamma^{-1}}\right) +
                    \frac{\int_{q_n}^{q_{n + 1}} f_H(h)dh}{q_n + \gamma^{-1}} = 0.
\end{align}
Multiplying the sides of~\eqref{eq:proof quantization levels part 2} by
$\frac{(q_n + \gamma^{-1})}{f_H(q_n)}$ we find~\eqref{eq:q_n equation
quantizer} and the theorem is proved.

\section{Proof of Theorem~\ref{prop:structure of uniform quantizer}}
\label{app:proof of prop structure of uniform quantizer}
Theorem~\ref{prop:structure of uniform quantizer} specifies the $N$
quantization levels for the uniform distribution.The proof follows
from~\eqref{eq:iterative formula for q_n}. We have
\begin{align}
(q_n + \gamma ^ {-1}) \left(1 +  \ln \left( \frac{q_n + \gamma^{-1}}
{q_{n - 1} + \gamma^{-1}} \right) \right)= q_{n + 1}+ \gamma^{-1}, \nonumber
\end{align}
which  leads to
\begin{align}
\label{eq:iterative formula for q_n 2}
\frac{q_{n + 1} + \gamma ^{-1}}{q_{n} + \gamma ^{-1}} =
            1 + \ln \left( \frac{q_{n} + \gamma ^{-1}}{q_{n - 1} + \gamma ^{-1}} \right).
\end{align}
Define $r_n = \frac{q_{n} + \gamma ^{-1}}{q_{n - 1} + \gamma ^{-1}}$.
Then we have $q_{n + 1} + \gamma^{-1} = r_n (q_n + \gamma^{-1})$ which
directly leads to
\begin{align}
q_n = \prod_{i = 0} ^ {n} r_i (q_{-1} + \gamma^{-1}),
\end{align}
where from $q_{-1} = 0$ in our earlier assumptions, we
find~\eqref{eq:q_n in terms of r_i}. Furthermore, by replacing the
ratios in~\eqref{eq:iterative formula for q_n 2} with $r_{n + 1}$ and
$r_n$ we find~\eqref{eq:iterative formula for r_i}. Finally,
from~\eqref{eq:finding_r_0} and~\eqref{eq:q_n in terms of r_i} we
obtain
\begin{align}
&(q_{N - 1} + \gamma ^{-1}) \ln(r_{N - 1}) = 2 - q_{N - 1} \nonumber \\
\Rightarrow &(q_{N - 1} + \gamma ^ {-1}) ( 1 + \ln(r_{N - 1})) = 2 + \gamma ^{-1} \nonumber \\
\Rightarrow &\frac{\prod_{n = 0} ^ {N - 1} r_n}{\gamma} r_N = 2 + \gamma ^{-1}.
\end{align}
which leads to~\eqref{eq:finding_r_0} and the theorem is proved. \hfill
$\blacksquare$

\section{Proof of Theorem~\ref{theorem: consistency}}
\label{app:proof of theorem consistency}

Theorem~\ref{theorem: consistency} sets the value of $\kappa_N$ to
ensure a consistent quantizer. To prove the theorem we need the
following lemma that sets an upper bound on an iterated logarithmic
sequence.
\begin{lemma}
\label{lemma:bound on r_i} For large $n$ and some positive constant
$c$, the iterated logarithmic sequence defined in~\eqref{eq:iterative
formula for r_i} is bounded as
\begin{align}
r_n \le 1 + \frac{c}{n}, \; n > 0.
\end{align}
\begin{proof}
Here we show that $r_n$ in~\eqref{eq:iterative formula for r_i}
decreases at least as $\mathcal{O}(\frac{1}{n})$ for large $n$. For
some $n > 1$ choose integer $M$ such that $r_n \le 1 + \frac{M}{n}$.
Then we have:
\begin{align}
r_{n + 1} = 1 + \ln r_n \le 1 + \ln\left(1 + \frac{M}{n}\right) =
                                        1 + \ln \left(\frac{M + n}{n}\right).
\end{align}
On the other hand, from the Reimann integral of the function $f(x) =
1/x$ over $[n, M + n)$ we have (choosing intervals of length $1$):
\begin{align}
\label{eq:app_r_n+1 from r_n}
\ln\left(\frac{M + n}{n}\right) &\le \sum_{k = n}^{M + n - 1} \frac{1}{k} =
                    \frac{1}{n} + \sum_{k = n + 1}^{n + M - 1} \frac{1}{k}  \nonumber \\
&\le \frac{1}{n} + \frac{(M - 1)}{n + 1} = \frac{M}{n + 1} + \frac{1}{n} - \frac{1}{n + 1} \nonumber \\
&\le \frac{M}{n + 1} + \mathcal{O}(\frac{1}{n^2}).
\end{align}
Then from induction and assuming a large enough $n$ we have, $r_{m} - 1
= \mathcal{O}(\frac{1}{m})$ for $m > n$.
\end{proof}
\end{lemma}
\vspace*{1ex} To prove Theorem~\ref{theorem: consistency}, we start
with the definition of $\delta(\mathbf{q})$
in~\eqref{eq:loss_general_eq_per_link}. It follows that
\begin{align}
\label{eq:proof_theorem_3_eq_1}
\delta(\mathbf{q}) &= \sum_{n = -1} ^ {N - 1} \int_{q_n}^{q_{n + 1}}
        \ln \left( \frac{h + \gamma ^{-1}}{q_n + \gamma^{-1}} \right) f_H(h) dh, \nonumber \\
&\le \sum_{n = -1} ^ {N - 2} \ln(r_{n + 1}) \left( F_H(q_{n + 1}) - F_H(q_{n}) \right)
                    + I_{N - 1} = \sum_{n = -1} ^ {N - 2} \ln(r_{n + 1}) Q_n + I_{N - 1},
\end{align}
where $Q_n = F_H(q_{n + 1}) - F_H(q_n)$ and $I_{N - 1}$ is defined
in~\eqref{eq:loss_general_eq_per_link}. We will argue that the sequence $\{Q_n\}_{n = 1} ^ {N - 1}$ is increasing with $n$. To see this, first note that the length of the quantization intervals increases with $n$ (we have $q_{n + 1} - q_n = q_n (r_{n + 1} - 1) \le q_n (1 - r_n^{-1}) = q_n - q_{n - 1}$ where the inequality is a direct result of the fact that for any $r> 1$, $r ^ {-1} + \ln r > 1$). Now from~\eqref{eq:finding_r_0_general} we have $q_{N - 1} \le \kappa_N$ and by fixing $\kappa_N = \kappa < N$ and using Lemma~\ref{lemma:bound on r_i} it is observed that $q_{n + 1} - q_n \le \kappa / N$. Therefore, by increasing $N$ the distance between quantization levels
approaches zero. This allows for Taylor series approximation of $F_H(h)$ at the quantization levels. Then, by assuming $f_H(h)$ to be almost constant over $[q_{n - 1}, q_{n + 1}]$, from the
definition of $Q_n$ for any $0 \le n < N - 1$ we have
\begin{align}
\label{:proof_theorem_3_eq_2}
\frac{Q_n}{Q_{n - 1}} &= \frac{F_H(q_{n + 1}) - F_H(q_n)}
                    {F_H(q_{n}) - F_H(q_{n - 1})} \nonumber \\
&\simeq \frac{f_H(q_n) (q_{n + 1} - q_n)}{f_H(q_{n}) (q_{n} - q_{n - 1})} =
                        \frac{q_n (r_{n + 1} - 1)}{q_n (1 - r_{n} ^ {-1})} > 1,
\end{align}
Note that for the uniform distribution the result above is valid for all $N > 1$. Considering the fact that $\sum_{n=-1}^{N-2} Q_n = F_H(q_{N-1})$, and $r_n$ being a strictly decreasing sequence (follows from the definition in~\eqref{eq:iterative formula for r_i}), it simply follows that
\begin{align}
\label{eq:proof_theorem_3_eq_2.5}
\sum_{n = -1} ^ {N - 2} Q_n \ln(r_{n + 1}) \le \sum_{n = -1} ^ {N - 2}
                                    \frac{F_H(q_{N - 1}) \ln(r_{n + 1})}{N - 1}.
\end{align}
By taking the natural logarithm of both sides
of~\eqref{eq:finding_r_0_general} we have
\begin{align}
\label{eq:proof_theorem_3_eq_3}
\sum_{n = -1} ^ {N - 2} \ln(r_{n + 1}) = \ln\left( \frac{\kappa \gamma + 1}{r_N} \right),
\end{align}
which together with~\eqref{eq:proof_theorem_3_eq_1}
and~\eqref{eq:proof_theorem_3_eq_2.5} leads to
\begin{align}
\label{eq:proof_theorem_3_eq_4}
\delta(\mathbf{q}) \le \frac{1}{N - 1} \ln\left(\frac{\kappa \gamma + 1}{r_N}\right) + I_{N - 1}.
\end{align}

In order to simplify~\eqref{eq:proof_theorem_3_eq_4}, we use
Lemma~\ref{lemma:bound on r_i} to find $r_N$ for $N \rightarrow
\infty$. Lemma~\ref{lemma:bound on r_i} shows that for $N\rightarrow
\infty$, $r_N \rightarrow 1$ leading to $q_{N - 1} \simeq \kappa$.
Therefore assuming $\gamma^{-1} \ll \kappa$, we can
write~\eqref{eq:proof_theorem_3_eq_4} in terms of $\kappa$ as in the
following
\begin{align}
\label{eq:proof_theorem_3_eq_5}
\delta(\mathbf{q}) \le \frac{\ln(\kappa\gamma )}{N} +
            \int_{\kappa}^{q_N} \ln\left( \frac{h}{\kappa} \right) f_H(h) dh.
\end{align}
The upper bound in~\eqref{eq:proof_theorem_3_eq_5} can be minimized
with respect to $\kappa$. This can be achieved by finding the root of
the derivative of~\eqref{eq:proof_theorem_3_eq_5}. Through some
cumbersome math it is found that the optimal $\kappa$, i.e., $\kappa^*$
satisfies
\begin{align}
\frac{1}{\kappa^* N} - \frac{1 - F_H(\kappa^*)}{\kappa^*} = 0 \nonumber \\
\Rightarrow \kappa^* = F_H^{-1}(1 - \frac{1}{N}).
\label{eq:proof_theorem_3_eq_6}
\end{align}
To obtain the smallest bound on $\delta(\mathbf{q})$, we need to
evaluate~\eqref{eq:proof_theorem_3_eq_5} at $\kappa = \kappa^*$. It
follows for $I_{N - 1}$ that
\begin{align}
\label{eq:proof_theorem_3_eq_7}
I_{N - 1} &\simeq \int_{\kappa^*}^{q_N} \ln \left( \frac{h}{\kappa^*}\right) f_H(h) dh \nonumber \\
          &= (1 - F_H(\kappa^*)) E\left[\ln \left( \frac{h}{\kappa^*}\right)
                                                \mid h > \kappa^* \right] \nonumber \\
          &\le \frac{1}{N} \ln \left( \frac{1}{\kappa^*}E[h \mid  h > \kappa^*]\right) \le
          \frac{c_1}{\kappa^*N},
\end{align}
where $c_1$ is a constant and the latest result is true due to the fact that
from~\eqref{eq:proof_theorem_3_eq_6}, $\kappa^* \rightarrow \infty$ as
$N \rightarrow \infty$. Finally, from~\eqref{eq:proof_theorem_3_eq_7}
and~\eqref{eq:proof_theorem_3_eq_5} we have
\begin{align}
\delta(\mathbf{q}) \le \mathcal{O} \left(\frac{\ln \kappa^*}{N} \right), \nonumber
\end{align}
and the theorem is proved. \hfill $\blacksquare$

\section{Proof of Theorem~\ref{theorem: high SNR}}
\label{app:proof of theorem high SNR} Theorem~\ref{theorem: high SNR}
states that the loss, is an order-$N$ iterated logarithm of SNR. To
prove the theorem, we will require the following lemma that proves some
useful properties of the ratios $r_n$;
\begin{lemma}
\label{lemma:properties of r_i} For all $n \ge 0$, the ratios $r_n$
defined according to~\eqref{eq:iterative formula for r_i} have the
following properties:
\begin{enumerate}
\item{$\lim_{\gamma \rightarrow \infty} r_n = \infty $}
\item{$\lim_{\gamma \rightarrow \infty} \frac{\gamma}{r_n} =
    \infty$}
\item{$\lim_{\gamma \rightarrow \infty} \frac{r_{n + 1}}{r_n} =
    \lim_{\gamma \rightarrow \infty} \frac{1 + \ln r_n}{r_n} = 0$}.
\end{enumerate}
\begin{proof}
The proof follows from~\eqref{eq:iterative formula for r_i}
and~\eqref{eq:finding_r_0_general}. Note that the left hand side
of~\eqref{eq:finding_r_0_general} is strictly increasing in $r_0$;
hence the first property. For the second property note that
from~\eqref{eq:iterative formula for r_i} we have $\gamma r_n ^ {-1} >
r_m \rightarrow \infty$ for $n \neq m$. Finally, the third property is
derived from~\eqref{eq:iterative formula for r_i} and the fact that
$\ln x/x \rightarrow 0$ for large $x$.
\end{proof}
\end{lemma}
\vspace*{1ex} To prove Theorem~\ref{theorem: high SNR}, consider a
quantizer with the ratios $r_n$ following~\eqref{eq:iterative formula
for r_i}. We proceed by finding an upper bound on the performance loss
$\delta(\mathbf{q})$ in~\eqref{eq:sum_In} in terms of the optimal
ratio's $r_n$ (defined in~\eqref{eq:iterative formula for r_i}). The
integrand in~\eqref{eq:sum_In} is increasing in $h$ and from the
definition of $r_n$ in Theorem~\ref{prop:structure of uniform
quantizer}, for any $n < N - 2$ we have
\begin{align}
\label{eq:I_n bound 1}
I_n \le \ln\left(r_{n + 1}\right) (F_H(q_{n + 1}) - F_H(q_n)),
\end{align}
where from~\eqref{eq:iterative formula for r_i}, $\ln r_{n - 1} = r_n -
1$. Letting $f_H^{\max}$ be the maximum value of $f_H(h)$, i.e.,
$f_H(H) \leq f_H^{\max}, \forall h > 0$, from~\eqref{eq:I_n bound 1}
and the definition of quantization levels in
Theorem~\ref{prop:quantization levels} it follows that
\begin{align}
\label{eq:I_n bound 2}
I_n \le (r_{n + 2} - 1) (q_{n + 1} - q_n) f_H^{max} \le \frac{(r_{n + 1} - 1) (r_{n + 2}- 1)
\prod_{m = 0} ^ {n} r_m }{\gamma} f_H^{max}
\end{align}
In order to find the limiting value of $I_n$ as $\gamma \rightarrow
\infty$ we use Lemma~\ref{lemma:properties of r_i}. From the first
property in Lemma~\ref{lemma:properties of r_i} and
equation~\eqref{eq:finding_r_0_general} we see that the upper bound
in~\eqref{eq:I_n bound 2} approaches zero as $\gamma \rightarrow
\infty$. Therefore, it follows for $-1 \le n < N - 2$ that
\begin{align}
\label{eq:high SNR I_n for n < N_2}
\lim_{\gamma \rightarrow \infty} I_n = 0.
\end{align}
For $I_{N - 2}$ we have
\begin{align}
I_{N - 2} \le \frac{(r_{N} - 1) (r_{N - 1}- 1) \prod_{m = 0} ^ {N - 2} r_m }{\gamma},
\end{align}
where from~\eqref{eq:finding_r_0_general} it readily follows that
\begin{align}
\label{eq:high SNR I_N_2}
\lim_{\gamma \rightarrow \infty} I_{N - 2} < \kappa_N < \infty.
\end{align}
Since in the high SNR regime $\gamma ^ {-1} \ll q_{N - 1}$,
from~\eqref{eq:sum_In} we have
\begin{align}
I_{N - 1} \simeq \int_{q_{N - 1}} ^ {\infty} \ln\left(\frac{h}{q_{N - 1}} \right) f_H(h) dh,
\end{align}
where due to concavity of the logarithm together with Jensen's inequality it follows that
\begin{align}
\label{eq:I_N-1 high SNR 1}
I_{N - 1} \le \ln\left( \frac{\bar{h}_{N-1}}{q_{N - 1}}\right) (1 - F_H(q_{N - 1})),
\end{align}
with $\bar{h}_{N-1} = E_h[h \mid h > q_{N - 1}]$.
From~\eqref{eq:finding_r_0_general} we have $q_{N - 1} =
\frac{\kappa_N}{r_N}$ and from the first property of
Lemma~\ref{lemma:properties of r_i} we see that $q_{N - 1} \rightarrow
0$ as $\gamma \rightarrow \infty$. This together with $E[h] = 1$ leads
to
\begin{align}
\label{eq:I_N-1 high SNR 2}
\bar{h}_{N-1} &= \int_{q_{N - 1}}^{\infty} \frac{h f_H(h)}{1 - F_H(q_{N - 1})} dh \nonumber \\
&\simeq (1 + F_H(q_{N - 1})) \int_{q_{N - 1}}^{\infty} h f_H(h) dh \nonumber \\
&\le (1 + q_{N - 1} f_H^{max}) E[h] \nonumber \\
&= 1 + q_{N - 1} f_H^{max}.
\end{align}
Finally from~\eqref{eq:I_N-1 high SNR 1} and~\eqref{eq:I_N-1 high SNR 2},
\begin{align}
\label{eq:high SNR I_N_1}
I_{N - 1} &\le \ln(f_H^{max} + \frac{1}{q_{N - 1}}) (1 - F_H(q_{N - 1})) \nonumber \\
          &\le \ln(f_H^{max}) + \ln \left(1 + \frac{r_N}{\kappa_N f_H^{max}}\right).
\end{align}
In conclusion, from~\eqref{eq:high SNR I_n for n < N_2},~\eqref{eq:high
SNR I_N_2}, and~\eqref{eq:high SNR I_N_1} we have
\begin{align}
\label{eq:high_snr_limit_loss}
\lim_{\gamma \rightarrow \infty} \delta(\mathbf{q}) &\le \lim_{\gamma \rightarrow \infty}
\left( \sum_{n = -1} ^ {N - 3} I_n + I_{N - 2} + I_{N - 1} \right) \nonumber \\
&\le 0 + \kappa_N + \ln(f_H^{max}) + \ln \left(1 + \frac{r_N}{\kappa_N f_H^{max}}\right)  \nonumber \\
&\sim \mathcal{O}\left(\ln^{(N+1)} r_0 \right) < \mathcal{O} \left( \ln^{(N + 1)} \gamma \right),
\end{align}
which completes the proof.

\bibliographystyle{IEEETran}
\bibliography{reflist}

\clearpage

\end{document}